\documentclass[12pt, draftcls, onecolumn]{IEEEtran}

\IEEEoverridecommandlockouts                              
\usepackage{graphicx}
\usepackage{graphics}
\usepackage{amsmath}
\usepackage{amssymb}
\usepackage{xy}
\usepackage{array}
\usepackage{cite}
\usepackage{setspace}
\usepackage{color}

\title{\LARGE
Design of Nonlinear State Observers for One-Sided Lipschitz Systems}

\date{}
\author{Masoud Abbaszadeh$^\dag$ \thanks{$\dag$ Author to whom correspondence should be addressed.},
    \and
    Horacio J. Marquez\\
     \normalsize{masoud@ualberta.net, marquez@ece.ualberta.ca\\
     \vspace{1.5mm} Department of Electrical
\& Computer Engineering, University of Alberta, Edmonton, Alberta, Canada, T6G 2V4}}

\begin{document}
\maketitle
\begin{abstract}
Control and state estimation of nonlinear systems satisfying a Lipschitz continuity condition have been important
topics in nonlinear system theory for over three decades, resulting in a substantial amount of literature.
The main criticism behind this approach, however, has been the restrictive nature of the Lipschitz continuity condition
and the conservativeness of the related results. This work deals with an extension to this problem by introducing a more general family of nonlinear functions, namely one-sided Lipschitz functions. The corresponding class of systems is a superset of its well-known Lipschitz counterpart and possesses inherent advantages with respect to conservativeness. In this paper, first the problem of state observer design for this class of systems is established, the challenges are discussed and some analysis-oriented tools are provided. Then, a solution to the observer design problem is proposed in terms of nonlinear matrix inequalities which in turn are converted into numerically efficiently solvable linear matrix inequalities.
\end{abstract}


\section{Introduction}

\PARstart{T}{he} observer design problem for nonlinear systems satisfying a Lipschitz continuity condition
has been a topic of a constant research for the last three decades. Observers
for Lipschitz systems were first considered by Thau in his seminal
paper \cite{Thau} where he obtained a sufficient condition to ensure
asymptotic stability of the observer error. Thau's condition provides a
useful analysis tool but does not address the fundamental
design problem. Inspired by Thau's work, several authors have studied
observer design for Lipschitz systems using various approaches
\cite{Rajamani,Aboky,Walcott,Zhu,Abbaszadeh4,Pertew2,Abbaszadeh5,Abbaszadeh1,Darouach3,Darouach4}.
Lipschitz systems constitute
an important class of nonlinear systems for which observer design can be carried out using pseudo-linear techniques.
Indeed, the Lipschitz continuity assumption provides a norm-based form of a nonlinear inequality which can be substituted into the observer error dynamics and, with some elaboration, the observer error dynamics can be cast in a numerically tractable format
that is dominated by a linear term.

The study of nonlinear Lipschitz systems is, however, overshadowed by some controversy surrounding the possibly restrictive
nature of the Lipschitz continuity condition. The statement is somewhat controversial and, in fact,
different researchers often refer to this condition as ``very general and satisfied by
most dynamical systems,'' and as ``too restrictive to be of any practical use.''  There is, of course, some truth in both of these
opposing views:  most nonlinear equations satisfy a Lipschitz continuity condition, with the same Lipschitz constant, only locally; possibly
in a small region of the state space. However, the Lipschitz condition is almost never satisfied globally thus placing a limit
to the application of any result based on a Lipschitz continuity condition.

The Lipschitz constant of such functions is usually region-based and often dramatically increases as the operating region is enlarged. On the other hand, even if the nonlinear system is Lipschitz in the region of interest, it is generally the case that the available
observer design techniques can only stabilize the error dynamics for dynamical systems with small Lipschitz constants
but, as discussed later,  fails to provide a solution when the Lipschitz constant becomes large. The problem becomes worse when dealing with \emph{stiff} systems. Stiffness means that the ordinary differential equation (ODE) admits a smooth solution with moderate derivatives, together with nonsmooth (``transient'') solutions rapidly converging towards the smooth ones \cite[p. 71]{Ency_math}. In particular, piece-wise functions such as saturation, deadband and backlash which are common in control systems (e.g. in actuators) are known to impose stiffness into the underlying system of ODEs. This problem has been recognized in the mathematical literature and specially in the field of numerical analysis for some time and a powerful tool has developed to overcome this problem. This tool is a generalization of the Lipschitz continuity to a less restrictive condition known as \emph{one-sided Lipschitz} continuity which has become one of the building blocks in numerical analysis and has been extensively applied to the stability analysis of ODE solvers \cite{Hairer2, Stuart, Dekker}.

Inspired by the these advances in the mathematical literature, in this paper, we extend this concept to the nonlinear observer design
problem and consider stabilization of the observer error dynamics based on the one-sided Lipschitz condition.
The advantages gained through this approach are two-fold: i) \emph{Generalization:} we will show that the one-sided Lipschitz continuity covers a broad family on nonlinear systems which includes the well-known Lipschitz systems as a special case. ii) \emph{Reduced conservativism:} Observer design techniques based on Lipschitz functions can guarantee stability only for small values of Lipschitz constants which directly translates into small stability regions. All available results on Lipschitz systems, however,
provide only sufficient conditions for stability and the actual observer might still work with larger Lipschitz constants,
even though the tool used in the analysis and design are unable to provide theoretical evidence.
The implication is that  there is a significant degree of conservativeness in the Lipschitz formulation, a critique that has often
been reported by researchers, but that has been difficult to correct and has produced no valuable alternative.

In this work we provide this valuable alternative in the form of the one-sided Lipschitz condition. We will show that the one-sided Lipschitz condition generalizes the classical Lipschitz theory in the following sense: any dynamical system satisfying a
Lipschitz condition satisfies also a one-sided Lipschitz condition. However,
the one-sided Lipschitz constant is always smaller than its Lipschitz counterpart,  a difference that can be significant even for very simple nonlinear functions \cite{Hairer2, Stuart, Dekker}. Examples are presented illustrating this property as well as showing cases where a dynamical system satisfies a one-sided Lipschitz condition even-though it is not Lipschitz in the classical sense.
Specially, when a dynamical system is \emph {stiff}, the conventional Lipschitz constant inevitably becomes very large while one-sided Lipschitz constant is still moderate \cite{Hairer2, Stuart, Dekker}. As a result, more efficient and less conservative observers can be developed in this context.

Unfortunately, these major advantages come along with a greater degree of difficulty encountered dealing with
one-sided Lipschitz systems. Unlike Lipschitz functions, which lead to an inequality in a rather simple quadratic form,
the one-sided Lipschitz formulation leads to a weighted bilinear form which imposes significant challenges in manipulating the Lyapunov derivative.

Very recently, nonlinear observers for one-sided Lipschitz systems were considered in \cite{Hu1, Hu2}.
These references consider a problem similar in spirit to Thau's original work on Lipschitz systems: given a
one-sided Lipschitz system with a known Lipschitz constant and a pre-defined Lyapunov function candidate,
obtain sufficient conditions for the existence of an observer. As with Thau's work, \cite{Hu1, Hu2} provide
an important analysis tool but the (difficult) design problem remains unsolved.

In this paper our goal is to acknowledge the advantages of the one-sided Lipschitz formulation over the conventional Lipschitz assumption in the control and observation theory, and in particular to formulate the observer design problem based on that. In this respect, not only do we provide basic analysis tools but also we address the \emph{design} problem
and present a complete solution. We introduce a new concept for nonlinear systems, the \emph{quadratic inner-boundedness}, which is discussed in Section IV. We show that the class of quadratic inner-bounded systems also extends the class of Lipschitz systems. Our observer design solution is based on the one-sided Lipschitz property and the quadratic inner-boundedness of the system.

The remainder of the paper is organized as follows:
Section II introduces the one-sided Lipschitz condition and study its basis properties. In Section III we consider the observer problem based on this property and addressed observer stability. Section IV, which contains the main results, addresses  observer design in the form of nonlinear matrix inequalities (NMIs). In the cycle, in order to use the efficient readily available numerical solvers, we convert the proposed NMI problem into linear matrix inequalities (LMIs). Section V presents an illustrative example.


\section{Mathematical Preliminaries and Problem Statement}

Throughout the paper
$\mathbb{R}$ represents the field of real numbers,
$\mathbb{R}^n$ the set of $n$-tuples of real numbers and $\mathbb{R}^{n \times p}$ the set of
real matrices of order $n$ by $p$.
$<,>$ is the (often called ``natural'') inner product in the space $\mathbb{R}^n$, {\it i.e.} given $x,y \in \mathbb{R}^n$, then
$<x,y>=x^{T}y$,
where $x^T$ is the transpose of the (column vector) $x \in \mathbb{R}^n$.
$\|.\|$ is the vector 2-norm (the Euclidian norm) in $\mathbb{R}^n$ defined by
$\|x\| = \sqrt{<x,x>}$.

Consider now the following continuous-time nonlinear dynamical system
\begin{align}
\dot{x}(t)&=Ax(t)+ \Phi(x,u)\hspace{7mm} A \in\mathbb{R}^{n\times
n}\label{con1}\\
y(t)&=Cx(t)\hspace{25mm} C \in\mathbb{R}^{n\times p},\label{con2}
\end{align}
where $x\in {\mathbb R} ^{n} ,u\in {\mathbb R} ^{m} ,y\in {\mathbb
R} ^{p} $ and $\Phi(x,u)$ represents a nonlinear function that is
continuous with respect to both $x$ and $u$. The system (\ref{con1})-\eqref{con2}
is said to be \emph{locally
Lipschitz} in a region $\mathcal{D}$ including the origin with
respect to $x$, uniformly in $u$, if there exist a constant $l>0$ satisfying:
\begin{eqnarray}
\|\Phi(x_{1},u^{*})-\Phi(x_{2},u^{*})\|\leqslant l\|x_{1}-x_{2}\|
\hspace{7mm}\forall \, x_{1} (t),x_{2} (t)\in \mathcal{D},\label{Lip}
\end{eqnarray}
where  $u^{*}$ is any admissible control signal. The smallest constant $l>0$ satisfying (\ref{Lip}) is known as the
\emph{Lipschitz constant}.
The region $\mathcal{D}$ is the \emph{operational region} or our \emph{region of interest}. If the condition
(\ref{Lip}) is valid everywhere in $\mathbb{R}^{n}$, then the function is said to be globally Lipschitz.

The following definition introduces one-sided Lipschitz functions.

\textbf{Definition 1. \cite{Hairer2}} The nonlinear function $\Phi(x,u)$ is said to be
\emph{one-sided Lipschitz} if there exist $\rho \in \mathbb{R}$ such that
\begin{eqnarray}
\left<\Phi(x_{1},u^{*})-\Phi(x_{2},u^{*}), x_{1}-x_{2}\right> \
\leqslant \rho \|x_{1}-x_{2}\|^{2} \hspace{7mm}\forall \, x_{1},x_{2}\in \mathcal{D},\label{con3}
\end{eqnarray}
where $\rho \in \mathbb{R}$ is called the \emph{one-sided Lipschitz constant}. As in the case of Lipschitz functions,
the smallest $\rho$ satisfying \eqref{con3} is called  the one-sided Lipschitz constant.\\

Similarly to the Lipschitz property, the one-sided Lipschitz property might be local or global.
Note that while the Lipschitz constant must be positive,
the one-sided Lipschitz constant can be positive, zero or even negative.
For any function $\Phi(x,u)$, we have:
\begin{align}
|\left<\Phi(x_{1},u^{*})-\Phi(x_{2},u^{*}), x_{1}-x_{2}\right>| &\leqslant \|\Phi(x_{1},u^{*})-\Phi(x_{2},u^{*})\|\|x_{1}-x_{2}\|\notag\\
\text{and if $\Phi(x,u)$ is Lipschitz, then:} \ \ \ \ \ &\leqslant  l \|x_{1}-x_{2}\|^{2}.
\end{align}
Therefore, any Lipschitz function is also one-sided Lipschitz. The converse, however, is not true.
For Lipschitz functions,
\begin{align}
-l \|x_{1}-x_{2}\|^{2} \leqslant \left<\Phi(x_{1},u^{*})-\Phi(x_{2},u^{*}), x_{1}-x_{2}\right> \
\leqslant l \|x_{1}-x_{2}\|^{2},
\end{align}
which is a \emph{two-sided} inequality v.s. the \emph{one-sided} inequality in \eqref{con3}.
If the nonlinear function $\Phi(x,u)$ satisfies the one-sided Lipschitz
continuity condition globally in $\mathbb{R}^{n}$, then the results are valid globally.
For continuously differentiable nonlinear functions it is well-known that the smallest possible
constant satisfying (\ref{Lip}) ({\it i.e.}, the Lipschitz constant) is the supremum of the norm of
Jacobian of the function over the region $\mathcal{D}$ (see for example \cite{Marquez}), that is:
\begin{align}
l= \limsup \left(\left\|\frac{\partial \Phi}{\partial x}\right\|\right), \ \ \ \ \forall x \in \mathcal{D}.
\end{align}
Alternatively, the one-sided Lipschitz constant is associated with the \emph{logarithmic matrix norm (matrix measure)} of the Jacobian.
The logarithmic matrix norm of a matrix $A$ is defined as \cite{Dekker}:
\begin{align}
\mu(A)= \lim_{\epsilon \rightarrow 0} \frac{|||I+\epsilon A |||-1}{\epsilon},\label{mu1}
\end{align}
where the symbol $|||.|||$ represents any matrix norm.
Then, we have \cite{Dekker}
\begin{align}
\rho= \limsup \left[\mu \left(\frac{\partial \Phi}{\partial x}\right)\right], \ \ \ \ \forall x \in \mathcal{D}.
\end{align}
If the norm used in \eqref{mu1} is indeed the induced 2-norm (the spectral norm) then it can be shown that $\mu(A)=\lambda_{max}\left(\frac{A+A^{T}}{2}\right)$ \cite{Vidyasagar1}.
On the other hand, from the Fan's theorem (see for example \cite{Horn1}) we know that for any matrix, $\lambda_{max}\left(\frac{A+A^{T}}{2}\right) \leq \sigma_{max}(A)=\|A\|$ \cite{Horn1}. Therefore $\rho \leq l$. Usually one-sided Lipschitz constant can be found to be much smaller than the Lipschitz constant \cite{Dekker}. Moreover, it is well-known in numerical analysis that for stiff ODE systems, $\rho << l$ \cite{Stuart, Dekker}. The one-sided Lipschitz continuity plays a vital role in solving such initial value problems. For instance, consider the following example which is taken from \cite[p. 173]{Stuart}.\\

\textbf{Example 1.} Suppose $\Phi(x)=-x^{3}$ with $x \in \mathbb{R}$. For the Lipschitz constant we have
\begin{align}
\|\Phi(x_{1})-\Phi(x_{2})\|=|x_{1}^{3}-x_{2}^{3}|\leq |x_{1}^{2}+x_{1}x_{2}+x_{2}^{2}|.|x_{1}-x_{2}|.\notag
\end{align}
By considering the line $x_{1}=x_{2}$, we deduce that the Lipschitz constant on any set
\begin{align}
\mathcal{D}=\{x \in \mathbb{R}: |x|\leq r\}, \notag
\end{align}
is necessarily $\geqslant 3r^{2}$. Therefore, this function is not globally Lipschitz but is locally Lipschitz with a region-based Lipschitz constant which dramatically increases with the region of interest. On the other hand, for the one-sided Lipschitz constant we have
\begin{align}
\left<\Phi(x_{1})-\Phi(x_{2}), x_{1}-x_{2}\right>&=-(x_{1}^{3}-x_{2}^{3})(x_{1}-x_{2})\notag\\
&=-(x_{1}-x_{2})^{2}(x_{1}^{2}+x_{1}x_{2}+x_{2}^{2})\leq -\frac{1}{2}(x_{1}-x_{2})^{2}(x_{1}^{2}+x_{2}^{2})\leq 0,\notag
\end{align}
which means that the function is globally one-sided Lipschitz with one-sided Lipschitz constant zero. We shall now illustrate an example which satisfies one-sided Lipschitz continuity but not Lipschitz continuity.\\

\textbf{Example 2.} A simple example of a one-sided Lipschitz function which indeed is not Lipschitz is $\Phi(x)=-sgn(x)\sqrt{|x|}$ with $x \in \mathbb{R}$, where $sgn(.)$ denotes the sign (signum) function. This function is monotone decreasing and so globally one-sided Lipschitz with one-sided Lipschitz constant  $\rho = 0$. In any interval $x \in [-m, m]$, the one-sided Lipschitz constant is $-\frac{1}{2\sqrt{m}}$ \cite{Donchev}. On the other hand, this function is not Lipschitz in any domain containing the origin. Indeed, for $x>0$,
\begin{align}
\frac{|f(x)-f(0)|}{|x-0|}=\frac{\sqrt{|x|}}{|x|}=\frac{1}{\sqrt{|x|}},
\end{align}
which is unbounded as $x\rightarrow 0$.\\

It is worth mentioning that the Lipschitz continuity property lies between continuity and continuous differentiability i.e. every Lipschitz function
is continuous but not necessarily continuously differentiable, while a one-sided Lipschitz function may be discontinuous. However, as in
the case of Lipschitz systems, one-sided Lipschitz systems have only one solution (if exists) associated with each initial condition. Therefore, if the
one-sided Lipschitz function is also continuous (a sufficient condition for the existence of a solution in ODEs), the system has a unique solution
for every initial condition \cite[p. 139]{Agarwal}. It can also be shown that if a dynamical system is locally Lipschitz and globally one-sided Lipschitz with a strictly negative one-sided Lipschitz constant, then the system has a unique exponentially attractive equilibrium point\cite{Stuart}. A comprehensive treatment of the interesting properties of the one-sided Lipschitz functions is beyond the scope of this paper. The interested reader can consult the literature on numerical analysis, such as \cite{Stuart}, \cite{Hairer2} and \cite{Dekker} and the references therein for further details.

\section{Observer Structure}

In this section we consider the observation problem; {\it i.e.} given the dynamical system
(\ref{con1})-(\ref{con2}) we assume that only the input $u$ and output $y$ are available and study the
feasibility of reconstructing the state $x$. To this end we consider an observer  of the following form
\begin{align}
\dot{\hat{x}}(t)=A\hat{x}(t)+\Phi(\hat{x},u)+L(y-C\hat{x})\label{observer1}.
\end{align}
The observer error dynamics is given by
\begin{align}
e(t)&\triangleq x(t)-\hat{x}(t),
\\\dot{e}(t)&=(A-LC)e(t)+\Phi-\hat{\Phi} \label{error1}
\end{align}
where $ \Phi\triangleq\Phi(x,u)$ and $  \hat{\Phi}\triangleq\Phi(\hat{x},u)$. Using a Lyapunov function candidate of the form $V(t)=e^{T}Pe$, the derivatives along the error trajectories are:
\begin{align}
\dot{V}=e^{T}[(A-LC)^TP+P(A-LC)]e+2e^{T}P(\Phi-\hat{\Phi})\label{Lyap1}.
\end{align}
Our goal is, assuming that $\Phi$ satisfies a one-sided Lipschitz condition,
find an observer gain $L$ such that the observer error dynamics is asymptotically stable.
Equation (\ref{Lyap1}), however, is independent of the assumption made on the function $\Phi$ and can
be used when $\Phi$ is either Lipschitz or one-sided Lipschitz.
Before further exploring the one-sided Lipschitz problem we digress to compare with the Lipschitz case.

\subsection{Motivating problem}
Given the form of equation $(\ref{Lyap1})$, the standard approach used in the Lipschitz case is to find $L$ such that
$(A-LC)^TP+P(A-LC) = -Q$ is \emph{negative definite} and $e^{T}Qe >  2e^{T}P(\Phi-\hat{\Phi})$. This approach,
first proposed by Thau in 1973 \cite{Thau}, has dominated the literature on Lipschitz systems ever since. The implicit
idea behind this approach is to use of the output injection term in the observer dynamics to ensure that the linear part of the
observer error dominates the nonlinear terms. This, in turn, is facilitated by the strong square norm condition
(\ref{Lip}) satisfied by Lipschitz systems which leads to the conservative nature of the result. In other words,
in the process of employing the Lipschitz property \eqref{Lip}, the term $e^{T}P(\Phi-\hat{\Phi})$ in the Lyapunov derivative
\eqref{Lyap1} is replaced by a strictly positive term forcing the rest of the
derivative to be sufficiently negative to compensate the remaining terms.

It is important to note that the term $e^{T}[(A-LC)^TP+P(A-LC)]e < 0$ if and only if $A-LC$ has eigenvalues with negative real part.
Unlike the Lipschitz constant $l$, which is positive by definition, the one-sided Lipschitz constant $\rho$
can be any real number. Thus,  the term $2e^{T}P(\Phi-\hat{\Phi})$ in \eqref{Lyap1} can be negative. Hence, a negative Lyapunov derivative may be guaranteed even with a positive definite  $(A-LC)^TP+P(A-LC)$ and consequently the linear
terms in $A-LC$ is not necessarily required to have eigenvalues with negative real part.
This means that the linear terms not necessarily dominate the nonlinear function $\Phi$, which in turn can lead to less conservative results.

The mathematical description behind the need of  $A-LC$ having eigenvalues in the left half plane can be traced back
through a substantial body of literature for Lipschitz systems such as \cite{Thau,Rajamani,Zhu,Abbaszadeh4,Pertew2,Abbaszadeh5}
while the freedom of the one-sided counterpart from such necessity is established in this article.

\subsection{Challenging Obstacle}
We now return to our main objective and endeavor to find $L$ that makes $\dot{V}<0$ in (\ref{Lyap1}).
This problem is nontrivial.  Unfortunately, to the best of the authors' knowledge, there is no result in mathematics
relating the weighted bilinear form $2e^{T}P(\Phi-\hat{\Phi})$ to the bilinear form  $2e^{T}(\Phi-\hat{\Phi})$
which is required to take full advantage of the properties offered by the one-sided Lipschitz formulation.
This is the main obstacle to overcome in this work.
A very simple first approach to this problem is to consider $P=I$. With this choice, from (\ref{Lyap1}) we have:
\begin{align}
\dot{V}=e^{T}[(A-LC)^T+(A-LC)]e+2e^{T}(\Phi-\hat{\Phi}) \leq e^{T}[(A-LC)^T+(A-LC)+2\rho I]e.\notag
\end{align}
where we substituted $e^T (\Phi - \hat{\Phi}) \leq \rho e^Te$. Hence, in order to have $\dot{V} < 0$, we must have
\begin{align}
(A-LC)^T+(A-LC)+2\rho I < 0 \Rightarrow \mu(A-LC) < -\rho.\label{Lyap3}
\end{align}
Inequality \eqref{Lyap3} is an LMI which can be efficiently solved using any available LMI solver to find the observer gain $L$. For the logarithmic matrix norm the following inequality can be used \cite{Vidyasagar1}:
\begin{align}
-\mu(-A) \leq \Re \lambda_{i}(A) \leq \mu(A), \ \ \ i=1,\ldots, n.
\end{align}
Therefore, $-\mu(-(A-LC)) \le \Re \lambda_{i}(A-LC)$. On the other hand, we want $\mu(A-LC) < -\rho$, so as a necessary condition, we must have
\begin{align}
\max_{i} \Re \lambda_{i}(A-LC)< -\rho.
\end{align}
Furthermore, suppose $A-LC$ is not stable (not stabilizable). We can always find $\alpha >0$ such that $(A-LC-\alpha I)$ is stable where $\alpha > \max_{i} \Re \lambda_{i}(A-LC)$. Then the observer error is asymptotically stable if
\begin{align}
(A-LC)^T+(A-LC)+2\rho I &= (A-LC-\alpha I)^T+(A-LC-\alpha I)+2\alpha I+2\rho I< 0\label{con5}
\end{align}
Now, a sufficient condition for \eqref{con5} to be true is
\begin{align}
\Rightarrow \alpha+\rho &<0 \Rightarrow \rho<-\alpha < - \max_{i} \Re \lambda_{i}(A-LC).\notag
\end{align}

\textbf{Remark 1.}
In general,  defining $P=I$ is too restrictive to be regarded as a viable solution as it forces the error dynamics to be stable with the Lyapunov candidate $V=e^{T}e$. Alternatively, we can redefine the one-sided Lipschitz property as
\begin{eqnarray}
\left<P\Phi(x_{1},u^{*})-P\Phi(x_{2},u^{*}), x_{1}-x_{2}\right> \
\leqslant \rho \|x_{1}-x_{2}\|^{2} \hspace{7mm}\forall \, x_{1}
(t),x_{2} (t)\in \mathcal{D}.\label{con4}
\end{eqnarray}
With this definition, we obtain $e^{T}P(\Phi-\hat{\Phi})\leq \rho e^{T}e$ and the resulting stability condition is
\begin{align}
(A-LC)^TP+P(A-LC)+2\rho I <0,\label{Lyap4}
\end{align}
which is an LMI. The idea is to find $\rho$, $P$ and $L$ such that \eqref{con4} and \eqref{Lyap4} are simultaneously satisfied. While \eqref{Lyap4} is a definite programming, \eqref{con4} is a function inequality that needs to be satisfied
for all points inside the region $\mathcal{D}$ for all times.

Solving the two inequalities together is also an open problem, much more difficult than finding $\rho$ in the standard one-sided definition \eqref{con3} which is still a challenging task. An alternative is to first find $P$ and $\rho$ from \eqref{con4} (which is also very difficult)
and then solve \eqref{Lyap4} for $L$. This approach, however, will not bring additional benefits as it is as restrictive as the previous solution in the sense that it forces the error dynamics to be stable with the \emph{specific} $P$ known \emph{apriori} from \eqref{con4}.
This is similar to the analysis result of \cite{Hu1, Hu2} in which $P$ is assumed to be known apriori.

The above discussion provides some analysis insight but does not address the fundamental design problem
in a satisfactory manner.  In the next section we propose a complete solution to this rather involved design problem.

\section{Main Results}

In this section, we first introduce the concept of \emph{quadratic inner-boundedness} for the function $\Phi(x,u)$.
Our design solution will make extensive use of this concept. \\

\textbf{Definition 2.} The nonlinear function $\Phi(x,u)$ is called \emph{quadratically inner-bounded} in the region $\mathcal{\tilde{D}}$ if $\forall \ x_{1}, x_{2} \in \widetilde{D}$ there exist $\beta, \gamma \in \mathbb{R}$ such that
\begin{align}
(\Phi(x_{1},u)-\Phi(x_{2},u))^{T}(\Phi(x_{1},u)-\Phi(x_{2},u))\leq \beta \|x_{1}-x_{2}\|^{2}+\gamma \left<x_{1}-x_{2},\Phi(x_{1},u)-\Phi(x_{2},u)\right>.\label{inner_bounded}
\end{align}

\begin{figure}[!h]
  \centering
  \includegraphics[width=4.5in]{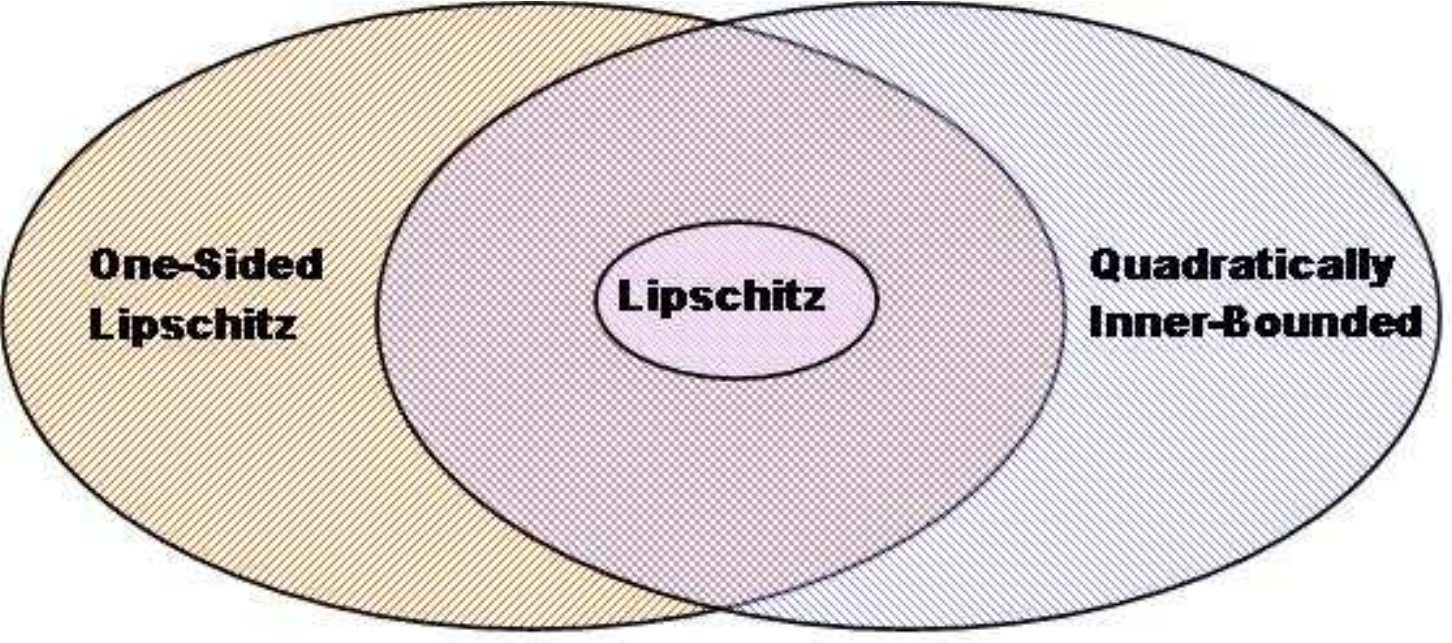}\\
  \caption{The Lipschitz, one-sided Lipschitz and quadratically inner-bounded function sets}\label{Fig1}
\end{figure}

It is clear that any Lipschitz function is also quadratically inner-bounded (e.g. with $\gamma=0$ and $\beta>0$). Thus, Lipschitz continuity implies quadratic inner-boundedness. The converse is, however, not true. We emphasize that $\gamma$ in \eqref{inner_bounded} can be any real number and is not necessarily positive. In fact, if $\gamma$ is restricted to be positive, then from the above definition, it can be concluded that $\Phi$ must be Lipschitz which is only a special case of our proposed class of systems. We will discuss the special Lipschitz cases in details, later in this section.
Figure \ref{Fig1} shows the relation between the Lipschitz, one-sided Lipschitz and quadratically inner-bounded function sets.

From now on we assume that $\Phi(x,u)$ is one-sided Lipschitz in $\mathcal{D}$ and quadratically inner-bounded in $\mathcal{\widetilde{D}}$.  All of our results will be valid in the intersection $\mathcal{D} \cap \mathcal{\widetilde{D}}$
(the operational region). If $\Phi(x,u)$ is both globally one-sided Lipschitz and globally quadratically inner-bounded, then all of our results will be valid globally. With the above notation, the following inequality holds for the estimation error.
\begin{align}
(\Phi-\hat{\Phi})^{T}(\Phi-\hat{\Phi})\leq \beta \|e\|^{2}+\gamma e^{T}(\Phi-\hat{\Phi}).
\end{align}

In the following Theorem, we propose a method for observer design for one-sided Lipschitz systems.

\textbf{Theorem 1.} Consider a nonlinear system satisfying inequalities \eqref{con3} and \eqref{inner_bounded} with constants $\rho$, $\beta$ and $\gamma$, along with the observer \eqref{observer1}. The observer error dynamics is asymptotically stable if there exist positive definite matrix $P$, symmetric matrix $Q$, matrix $L$ and a positive scalar $\alpha>0$ such that the following matrix inequalities problem is feasible:
\begin{align}
&(A-LC)^{T}P+P(A-LC)\leq -Q, \label{NMI1}\\
&\xi \lambda_{max}(P)-\lambda_{min}(P)< \alpha \lambda_{min}(Q), \label{NMI2}\\
&\gamma+2\alpha>0,\label{NMI3}\\
&\frac{\lambda_{max}(P)}{\lambda_{min}(P)}.(\alpha^{2}-1)< \alpha^{2},\label{kappa}
\end{align}
where $\xi=(\beta+1)+\rho(\gamma+2\alpha)$.\\

\textbf{Proof:} We have
\begin{align}
\dot{V}=e^{T}[(A-LC)^TP+P(A-LC)]e+2e^{T}P(\Phi-\hat{\Phi}).
\end{align}
for any nonzero $\alpha$ we can write
\begin{align}
2e^{T}P(\Phi-\hat{\Phi})&=\frac{1}{\alpha}\left[e+\alpha(\Phi-\hat{\Phi})\right]^{T}P\left[e+\alpha(\Phi-\hat{\Phi})\right]\notag\\
&-\frac{1}{\alpha}e^{T}Pe-\alpha (\Phi-\hat{\Phi})^{T}P(\Phi-\hat{\Phi}), \ \  \  \forall \alpha \in \mathbb{R}-\{0\}.
\end{align}
Assuming $\alpha>0$, then
\begin{align}
2e^{T}P(\Phi-\hat{\Phi}) \leq \frac{1}{\alpha}\lambda_{max}(P)\left[e+\alpha(\Phi-\hat{\Phi})\right]^{T}\left[e+\alpha(\Phi-\hat{\Phi})\right]
-\frac{1}{\alpha}e^{T}Pe-\alpha (\Phi-\hat{\Phi})^{T}P(\Phi-\hat{\Phi})\label{ineqII}.
\end{align}
Using the quadratic inner-boundedness property we have
\begin{align}
&e^{T}e+\alpha^{2}(\Phi-\hat{\Phi})^{T}(\Phi-\hat{\Phi})\leq (\beta+1) e^{T}e
+\gamma e^{T}(\Phi-\hat{\Phi})+(\alpha^{2}-1)(\Phi-\hat{\Phi})^{T}(\Phi-\hat{\Phi})\notag\\
&\Rightarrow \left[e+\alpha (\Phi-\hat{\Phi})\right]^{T}\left[e+\alpha (\Phi-\hat{\Phi})\right] \leq (\beta+1)e^{T}e+(\gamma+2\alpha)e^{T}(\Phi-\hat{\Phi})+(\alpha^{2}-1)\|\Phi-\hat{\Phi}\|^{2}.\label{ineqV}
\end{align}
Substituting \eqref{ineqV} into \eqref{ineqII} leads to
\begin{align}
2e^{T}P(\Phi-\hat{\Phi}) \leq &\frac{1}{\alpha}\lambda_{max}(P)\left[(\beta+1)e^{T}e+(\gamma+2\alpha)e^{T}(\Phi-\hat{\Phi})+(\alpha^{2}-1)\|\Phi-\hat{\Phi}\|^{2}\right]\notag\\
&-\frac{1}{\alpha}e^{T}Pe-\alpha (\Phi-\hat{\Phi})^{T}P(\Phi-\hat{\Phi})\label{ineq1}.
\end{align}
Based on the Rayleigh's inequality, for any $\alpha>0$ we have
\begin{align}
(\Phi-\hat{\Phi})^{T}P(\Phi-\hat{\Phi})\geq \lambda_{min}(P)\|\Phi-\hat{\Phi}\|^{2} \Rightarrow -\alpha(\Phi-\hat{\Phi})^{T}P(\Phi-\hat{\Phi})\leq -\alpha \lambda_{min}(P)\|\Phi-\hat{\Phi}\|^{2}\notag.
\end{align}
Hence, from \eqref{ineq1}, using the one-sided Lipschitz inequality \eqref{con3} and knowing that $\gamma+2\alpha>0$, we obtain\begin{align}
2e^{T}P(\Phi-\hat{\Phi}) \leq& \left[\frac{1}{\alpha}\lambda_{max}(P)(\alpha^{2}-1)-\alpha\lambda_{min}(P)\right]\|\Phi-\hat{\Phi}\|^{2}\notag\\
&+\frac{1}{\alpha}\lambda_{max}(P)\left[(\beta+1)+\rho(\gamma+2\alpha)\right]e^{T}e-\frac{1}{\alpha}e^{T}Pe.
\end{align}
We know that $\kappa(P)(\alpha^{2}-1)< \alpha^{2}$, where $\kappa(P)$ is the condition number of $P$ or $\frac{1}{\alpha}\lambda_{max}(P)(\alpha^{2}-1)-\alpha\lambda_{min}(P)< 0$. Then,
\begin{align}
2e^{T}P(\Phi-\hat{\Phi}) < \frac{1}{\alpha}\lambda_{max}(P)\left[(\beta+1)+\rho(\gamma+2\alpha)\right]e^{T}e-\frac{1}{\alpha}e^{T}Pe.\label{ineqVI}
\end{align}
Now we substitute \eqref{ineqVI} into the Lyapunov derivative. We obtain
\begin{align}
\dot{V}&=e^{T}[(A-LC)^{T}P+P(A-LC)]e+2e^{T}P(\Phi-\hat{\Phi}) \notag \\
& \leq e^{T}\left[(A-LC)^{T}P+P(A-LC)\right]e+\frac{\xi}{\alpha}\lambda_{max}(P)e^{T}e-\frac{1}{\alpha}e^{T}Pe,\label{ineq2}
\end{align}
where $\xi\triangleq(\beta+1)+\rho(\gamma+2\alpha)$. Therefore, in order to have $\dot{V}< 0$ it suffices to have:
\begin{align}
&(A-LC)^{T}P+P(A-LC)+\frac{\xi}{\alpha}\lambda_{max}(P)I-\frac{1}{\alpha}P < 0 \notag\\
&\Leftrightarrow \frac{\xi}{\alpha}\lambda_{max}(P)+\lambda_{max}\left[(A-LC)^{T}P+P(A-LC)-\frac{1}{\alpha}P\right]< 0\label{ineqVII}.
\end{align}
For any two symmetric matrices $A$ and $B$, it can be shown that $\lambda_{i}\leq \lambda_{i}(A)+\lambda_{i}(B)$, where $\lambda_{i}$s are the sorted eigenvalues\cite{Horn1}. Thus,
\begin{align}
\lambda_{max}&\left[(A-LC)^{T}P+P(A-LC)-\frac{1}{\alpha}P\right]\leq \lambda_{max}\left[(A-LC)^{T}P+P(A-LC)\right]+\lambda_{max}\left(-\frac{P}{\alpha}\right)\notag\\
&=\lambda_{max}\left[(A-LC)^{T}P+P(A-LC)\right]-\frac{1}{\alpha}\lambda_{min}(P)\label{ineqVIII}.
\end{align}
Now without loss of generality, we assume that there exists a symmetric matrix $Q$, such that $(A-LC)^{T}P+P(A-LC)\leq -Q$. Note that $Q$
is not necessarily positive definite (meaning that $(A-LC)$ is not necessarily stable). Thus,
\begin{align}
\lambda_{max}\left[(A-LC)^{T}P+P(A-LC)\right]\leq \lambda_{max}(-Q)=-\lambda_{min}(Q).\label{ineq3}
\end{align}
Substituting from \eqref{ineq3}, \eqref{ineqVIII} and \eqref{ineqVII} into \eqref{ineq2}, we get
\begin{align}
\frac{\xi}{\alpha}\lambda_{max}(P)-\frac{1}{\alpha}\lambda_{min}(P)-\lambda_{min}(Q)<0.\label{ineq4}
\end{align}
This completes the proof. $\blacksquare$\\

\textbf{Remark 2.} From a second glance at \eqref{kappa}, it is clear that if $\alpha^{2}-1\leq 0$ ($0<\alpha\leq 1$, taking $\alpha>0$ into consideration), then the inequality is always satisfied. On the other hand, if $\alpha^{2}-1>0$ ($\alpha>1$, knowing that $\alpha>0$), we have $\kappa(P)< \frac{\alpha^{2}}{\alpha^{2}-1}$. Then, as $\alpha$ approached one, the admissible condition number of $P$ becomes larger. If $\gamma > -2$, we can always pick $\alpha = 1$ so that both \eqref{NMI3} and \eqref{kappa} are always satisfied. On the other hand, if $\xi \leq 0$, then \eqref{NMI2} is always satisfied. So, if in addition, $\alpha$ can be chosen such that $\xi \leq 0$, then the NMIs in Theorem 1 reduce to a simple LMI feasibility problem (solving only \eqref{NMI1}).

\subsection{Special Lipschitz cases}
As mentioned earlier, the class of nonlinear systems we are addressing here covers Lipschitz systems as a special case. It is worth to briefly study under what conditions one-sided Lipschitz systems reduce to the Lipschitz case.

\begin{itemize}
  \item \textbf{Case 1.} If \eqref{inner_bounded} is satisfied with $\gamma=0$, then $\beta$ has to be positive and $\Phi(x,t)$ is Lipschitz with Lipschitz constant $\sqrt{\beta}$.
  \item \textbf{Case 2.} If for a one-sided Lipschitz nonlinear function $\Phi(x,u)$, \eqref{inner_bounded} is satisfied with $\gamma>0$, then $\beta+\gamma\rho$ has to be positive and it is easy to show that $\Phi(x,u)$ is Lipschitz with Lipschitz constant $\sqrt{\beta+\gamma\rho}$.
  \item \textbf{Case 3.} As mentioned in Remark 2, if $\alpha^{2}-1\leq 0$ ($0<\alpha\leq 1$), then the inequality \eqref{kappa} is always satisfied. On the other hand, from \eqref{ineqV} and knowing that $\alpha^{2}-1\leq 0$ and $\gamma+2\alpha>0$, we can write
\begin{align}
&\left[e+\alpha (\Phi-\hat{\Phi})\right]^{T}\left[e+\alpha (\Phi-\hat{\Phi})\right] \leq (\beta+1)e^{T}e+(\gamma+2\alpha)e^{T}(\Phi-\hat{\Phi})\notag\\
&\leq [(\beta+1)+(\gamma+2\alpha)\rho]e^{T}e \Rightarrow \|e+\alpha (\Phi-\hat{\Phi})\|^{2} \leq [(\beta+1)+(\gamma+2\alpha)\rho]\|e\|^{2}.
\end{align}
\end{itemize}
As a consequence, we have
\begin{align}
&-\|e\|+\alpha\|\Phi-\hat{\Phi}\|\leq \|e+\alpha(\Phi-\hat{\Phi})\| \leq \sqrt{(\beta+1)+(\gamma+2\alpha)\rho}\ \|e\|\notag\\
&\Rightarrow \|\Phi-\hat{\Phi}\|\leq \frac{1}{\alpha}\left[1+\sqrt{(\beta+1)+(\gamma+2\alpha)\rho}\right]\|e\|,
\end{align}
which means that $\Phi$ is Lipschitz. In other words, only for Lipschitz functions we can find $0<\alpha \leq 1$ such that $\gamma+2\alpha>0$ and \eqref{ineqVI} are both satisfied.

\subsection{LMI formulation}

Theorem 1, provides a design method for nonlinear observers for one-sided Lipschitz systems in the form of the
nonlinear matrix inequalities (NMIs) \eqref{NMI1}-\eqref{kappa}.
The difficulty, however, is that although Theorem 1 provides a legitimate solution to our problem,
there is currently no efficient solution in the numerical analysis literature capable of solving NMIs.
Unlike the nonlinear case, however, \emph{linear} matrix inequalities (LMIs) can be efficiently solved using commercially available packages such as the Matlab LMI solver. We now show how to cast the proposed nonlinear matrix inequalities solution into the LMI framework to take advantage of the efficient numerical LMI solvers readily available.

Using Fan's theorem mentioned earlier, we can write
\begin{align}
\lambda_{max}[(A-LC)^{T}P+P(A-LC)] \leq 2\sigma_{max}[P(A-LC)] \leq 2\lambda_{max}(P)\sigma_{max}(A-LC).
\end{align}
Substituting this back to \eqref{ineqVII} yields to
\begin{align}
&\frac{1}{\alpha}\lambda_{max}(P)\xi -\frac{1}{\alpha}\lambda_{min}(P)+2\lambda_{max}(P)\sigma_{max}(A-LC) < 0 \notag\\
&\Rightarrow \lambda_{max}(P)\left[2\sigma_{max}(A-LC)+\frac{\xi}{\alpha}\right]< \frac{1}{\alpha}\lambda_{min}(P) \Rightarrow \kappa(P)\left[2\sigma_{max}(A-LC)+\frac{\xi}{\alpha}\right]< \frac{1}{\alpha}\notag \\
& \Rightarrow \sigma_{max}(A-LC)< \frac{1}{2\alpha}\left(\frac{1}{\kappa(P)}-\xi\right),
\end{align}
which by means of Schur's complement and change of variable $\lambda=\frac{1}{\kappa}$ is equivalent to the LMI \eqref{LMI1}. LMI \eqref{LMI2} represents the condition $\kappa(\alpha^{2}-1)<\alpha^{2}$.

Based on the above discussion which also serves as the proof, the following corollary provides an LMI solution to our observer design problem. \\

\textbf{Corollary 1.} Consider a nonlinear system satisfying inequalities \eqref{con3} and \eqref{inner_bounded} with constants $\rho$, $\beta$ and $\gamma$, along with the observer \eqref{observer1}. The observer error dynamics is asymptotically stable if there exists a matrix $L$ and positive scalars $\alpha>0$ and $0<\lambda<1$ such that the following matrix inequalities problem is feasible:
\begin{align}
&\left[
  \begin{array}{cc}
    \frac{1}{2\alpha}\left(\lambda-\xi\right)I & (A-LC)^{T}\\
    (A-LC) & \frac{1}{2\alpha}\left(\lambda-\xi\right)I \\
  \end{array}
\right]>0,\label{LMI1}\\
&\gamma+2\alpha>0,\\
&\lambda > 1-\frac{1}{\alpha^{2}},\label{LMI2}
\end{align}
where $\xi=(\beta+1)+\rho(\gamma+2\alpha)$.\\

We now summarize the observer design procedure based on Corollary 1.

\subsection{Design Procedure}

\begin{itemize}
  \item \textbf{Step 1:} Pick an $\alpha>0$ such that $2\alpha+\gamma>0$.
  \item \textbf{Step 2:} If $\rho=0$ and $\beta \ge 0$, Stop; otherwise, calculate $\xi=\beta+\rho(\gamma+2\alpha)+1$.
  \item \textbf{Step 3:} Check if the conditions $\lambda-\xi>0$ and $0 < \lambda < 1$ are consistent. If Yes, go to Step 4;
                         otherwise, adjust $\alpha$ (if $\rho>0$ then decrease $\alpha$, otherwise increase $\alpha$) and go to Step 1.
  \item \textbf{Step 4:} Solve the LMIs in Corollary 1 for $L$ and $\lambda$.\\
\end{itemize}
Note that if Step 3 is passed, the LMIs in Step 4 are always feasible meaning that an observer gain $L$ will always be found.
In order for a feasible solution to exist, we require both $2\alpha+\gamma>0$ and $\xi<1$ to be satisfied. Therefore, one can choose $\alpha>0$ such that $2\alpha+\gamma>0$ and $\beta+\rho(\gamma+2\alpha)<0$ are simultaneously satisfied. This guarantees the feasibility of the solution space in Corollary 1. A structural limitation here is that if $\rho=0$, then $\beta$ has to be negative. The variable $\kappa (=\frac{1}{\lambda})$ calculated in Corollary 1 is the condition number of the $P$ matrix used in the Lyapunov function. Any $P$ with such condition number would be acceptable. Although easy to do, as our goal of finding the observer gain $L$ is already achieved, this step (finding $P$) is unnecessary except for analysis purposes of the results.\\

\textbf{Remark 3.} The beauty of the above result is in its simplicity and in having the LMI form. It is, however, worth mentioning that this solution is conservative in the sense that we have replaced $\lambda_{max}[(A-LC)^{T}P+P(A-LC)]$ which may or may not be positive, with the positive quantity $\lambda_{max}(P)\sigma_{max}(A-LC)$. Nonetheless, the solution provides a major improvement over the original solution in Theorem 1, in terms of numerical solvability.\\

\textbf{Remark 4.} There is yet another advantage in the LMI formulation. The one-sided Lipschitz constant $\rho$ appears linearly in the proposed LMIs. Hence, one can take advantage of the convexity of the solution space of the LMIs and solve an LMI optimization problem, maximizing the admissible one-sided Lipschitz constant for which the observer error stability is guaranteed. The resulting maximized $\rho$ acts as an upper bound to those of nonlinear functions $\Phi(s)$. This is specially helpful in the situations that an accurate estimate of the one-sided Lipschitz constant is hard to obtain. Alternatively, the optimization can be performed over the quadratic inner-boundedness parameters, $\beta$ or $\gamma$.

\section{Illustrative Example}
In this section, we illustrate the proposed observer design procedure through a numerical example.

\begin{figure}[!htb]
  \centering
  \includegraphics[width=6.5in]{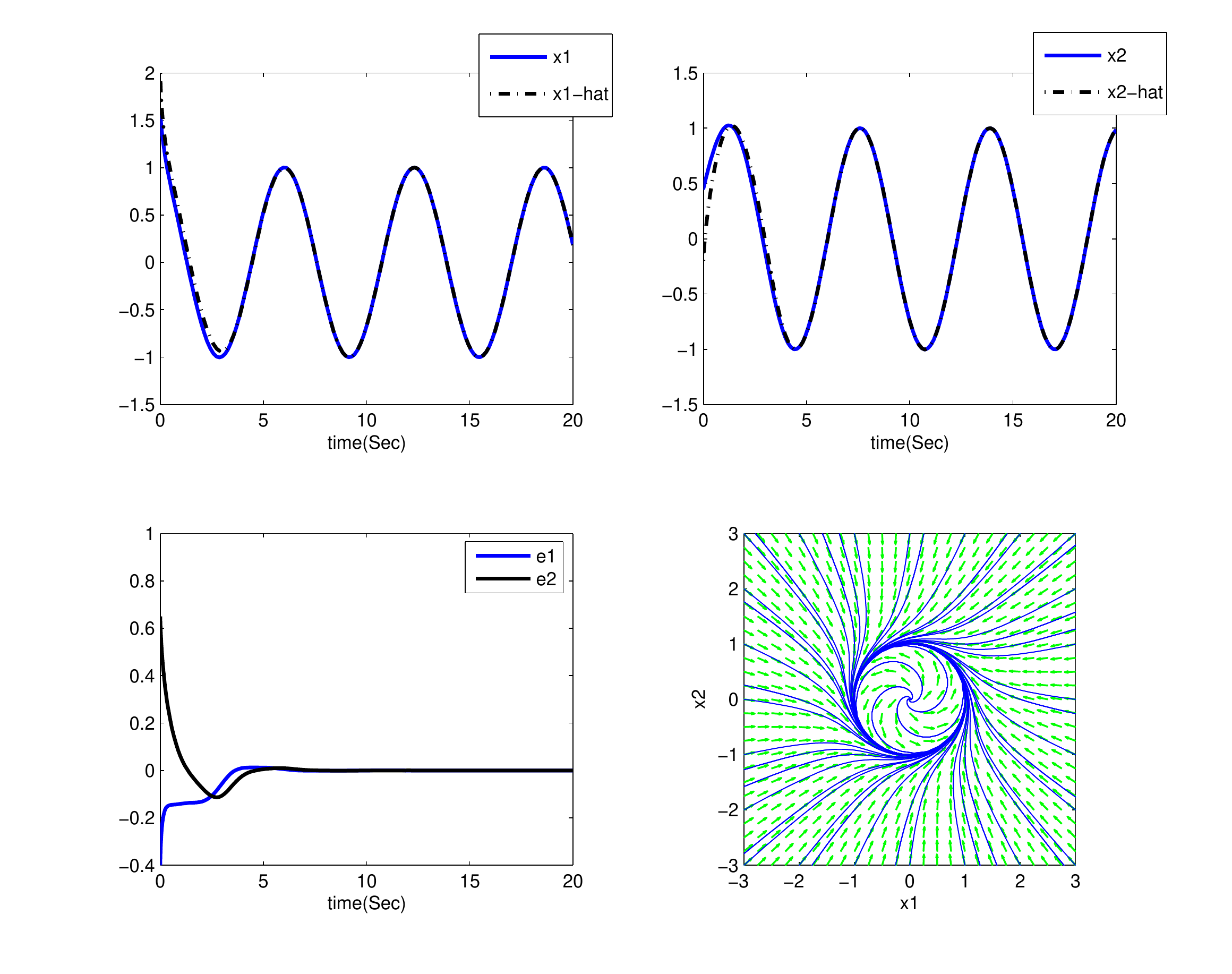}\\
  \caption{The system and observer states and the phase plane}\label{Fig2}
\end{figure}

\textbf{Example 3.} Suppose that the equations of motion of a moving object are given in the 2D Cartesian coordinates as follows:
\begin{align}
&\dot{x}=x-y-x(x^{2}+y^{2}),\notag\\
&\dot{y}=x+y-y(x^{2}+y^{2}),\notag
\end{align}
and $y$ is measured. We define the state vector as $\textbf{x}=\left[
                                                                   \begin{array}{cc}
                                                                     x & y \\
                                                                   \end{array}
                                                                 \right]^{T}$ and $\textbf{y} = y$ as the output, in which the variables are bolded to
avoid ambiguity. We have:
\begin{align}
&\dot{\textbf{x}}=\left[\begin{array}{cc}
                   1 & -1 \\
                   1 & 1
                 \end{array}
                 \right]\textbf{x}+\left[
                                     \begin{array}{c}
                                       -x(x^{2}+y^{2}) \\
                                       -y(x^{2}+y^{2}) \\
                                     \end{array}
                                   \right],\notag\\
&\textbf{y}=\left[
             \begin{array}{cc}
               0 & 1 \\
             \end{array}
           \right]\textbf{x}.\notag
\end{align}
Then we verify the one-sided Lipschitz property.
\begin{align}
\langle\Phi(\textbf{x}_{1})-\Phi(\textbf{x}_{2}), \textbf{x}_{1}-\textbf{x}_{2}\rangle=(x_{1}x_{2}+y_{1}y_{2})\left[(x_{1}^{2}+y_{1}^{2})+(x_{2}^{2}+y_{2}^{2})\right]
-(x_{1}^{2}+y_{1}^{2})^{2}-(x_{2}^{2}+y_{2}^{2})^{2}.\notag
\end{align}
On the other hand, $x_{1}x_{2}=\frac{1}{2}(x_{1}^{2}+x_{2}^{2})-\frac{1}{2}(x_{1}-x_{2})^{2}$. Thus,
\begin{align}
x_{1}x_{2}+y_{1}y_{2}=\frac{1}{2}\left[\|\textbf{x}_{1}\|^{2}+\|\textbf{x}_{2}\|^{2}-\|\textbf{x}_{1}-\textbf{x}_{2}\|^{2}\right].\label{examp1}
\end{align}
Therefore,
\begin{align}
\langle\Phi(\textbf{x}_{1})-\Phi(\textbf{x}_{2}), \textbf{x}_{1}-\textbf{x}_{2}\rangle&=\frac{1}{2}\left[\|\textbf{x}_{1}\|^{2}+\|\textbf{x}_{2}\|^{2}-\|\textbf{x}_{1}-\textbf{x}_{2}\|^{2}\right].
\left[\|\textbf{x}_{1}\|^{2}+\|\textbf{x}_{2}\|^{2}\right]-\|\textbf{x}_{1}\|^{4}-\|\textbf{x}_{2}\|^{4}\notag\\
&=-\frac{1}{2}\|\textbf{x}_{1}-\textbf{x}_{2}\|^{2}\left[\|\textbf{x}_{1}\|^{2}+\|\textbf{x}_{2}\|^{2}\right]
-\frac{1}{2}\left[\|\textbf{x}_{1}\|^{2}-\|\textbf{x}_{2}\|^{2}\right]^{2}\notag\\
&\leq 0.\label{inner}
\end{align}
This means that the systems is globally one-sided Lipschitz with the one-sided Lipschitz constant $\rho=0$. Now, lets verify the Lipschitz continuity property. $\Phi$ is continuously differentiable, so an estimate for the Lipschitz constant is the supremum of the norm of the Jacobian matrix, $J$.
We have:
\begin{align}
J=\left[
    \begin{array}{cc}
      -3x^{2}-y^{2} & -2xy \\
      -2xy & -3y^{2}-x^{2} \\
    \end{array}
  \right].\notag
\end{align}
$J$ is a symmetric matrix; therefore, its induced 2-norm equals its spectral radius.
\begin{align}
\|J\|=\sigma_{max}(J)=\max_{i}|\lambda_{i}(J)|=\max(|-2(x^{2}+y^{2})\pm (x^{2}+y^{2})|)=3(x^{2}+y^{2})=3r^{2}.\notag
\end{align}
This means that the system is locally Lipschitz and on any set
\begin{align}
\mathcal{D}=\{\textbf{x} \in \mathbb{R}^{2}: \|\textbf{x}\|\leq r\}, \notag
\end{align}
the Lipschitz constant $l$ is $3r^{2}$, i.e. the Lipschitz constant rapidly increases with the increase of $r$. We will come back to this later. Now we proceed our one-sided Lipschitz based observer design. We need to verify the quadratic inner-boundedness property of the system, as well. The left hand side of the quadratic inner-boundedness \eqref{inner_bounded} is:
\begin{align}
\textrm{LHS}&: \left[\Phi(\textbf{x}_{1})-\Phi(\textbf{x}_{2})\right]^{T}\left[\Phi(\textbf{x}_{1})-\Phi(\textbf{x}_{2})\right]\notag\\
&=(x_{1}^{2}+y_{1}^{2})^{3}+(x_{2}^{2}+y_{2}^{2})^{3}-2(x_{1}x_{2}+y_{1}y_{2})(x_{1}^{2}+y_{1}^{2})(x_{2}^{2}+y_{2}^{2})\notag\\
&=\|\textbf{x}_{1}\|^{6}+\|\textbf{x}_{2}\|^{6}-\left[\|\textbf{x}_{1}\|^{2}+
\|\textbf{x}_{2}\|^{2}-\|\textbf{x}_{1}-\textbf{x}_{2}\|^{2}\right]\|\textbf{x}_{1}\|^{2}\|\textbf{x}_{2}\|^{2}\notag\\
&=\|\textbf{x}_{1}\|^{4}\left[\|\textbf{x}_{1}\|^{2}-\|\textbf{x}_{2}\|^{2}\right]+\|\textbf{x}_{2}\|^{4}\left[\|\textbf{x}_{2}\|^{2}-\|\textbf{x}_{1}\|^{2}\right]
+\|\textbf{x}_{1}-\textbf{x}_{2}\|^{2}\|\textbf{x}_{1}\|^{2}\|\textbf{x}_{2}\|^{2},\notag\\
&=\left[\|\textbf{x}_{1}\|^{2}-\|\textbf{x}_{2}\|^{2}\right]^{2}.\left[\|\textbf{x}_{1}\|^{2}+\|\textbf{x}_{2}\|^{2}\right]
+\|\textbf{x}_{1}-\textbf{x}_{2}\|^{2}\|\textbf{x}_{1}\|^{2}\|\textbf{x}_{2}\|^{2},\notag
\end{align}
in which \eqref{examp1} is used. The right hand side of the quadratic inner-boundedness \eqref{inner_bounded} is:
\begin{align}
\textrm{RHS}&: \gamma \langle\Phi(\textbf{x}_{1})-\Phi(\textbf{x}_{2}), \textbf{x}_{1}-\textbf{x}_{2}\rangle+\beta\|\textbf{x}_{1}-\textbf{x}_{2}\|^{2}\notag\\
&=-\frac{\gamma}{2}\|\textbf{x}_{1}-\textbf{x}_{2}\|^{2}\left[\|\textbf{x}_{1}\|^{2}+\|\textbf{x}_{2}\|^{2}\right]
-\frac{\gamma}{2}\left[\|\textbf{x}_{1}\|^{2}-\|\textbf{x}_{2}\|^{2}\right]^{2}+\beta\|\textbf{x}_{1}-\textbf{x}_{2}\|^{2}\notag\\
&=\|\textbf{x}_{1}-\textbf{x}_{2}\|^{2}\left[\beta-\frac{\gamma}{2}\left(\|\textbf{x}_{1}\|^{2}+\|\textbf{x}_{2}\|^{2}\right)\right]
-\frac{\gamma}{2}\left[\|\textbf{x}_{1}\|^{2}-\|\textbf{x}_{2}\|^{2}\right]^{2},\notag
\end{align}
in which \eqref{inner} is used. We have to find values for $\beta$ and $\gamma$ and a region $\mathcal{\widetilde{D}}$ such that for all $\textbf{x}\in \mathcal{\widetilde{D}}$, $\textrm{LHS} \leq \textrm{RHS}$. Comparing the two, it suffices to have:
\begin{align}
&\|\textbf{x}_{1}\|^{2}+\|\textbf{x}_{2}\|^{2} \leq -\frac{\gamma}{2},\notag\\
&\|\textbf{x}_{1}\|^{2}.\|\textbf{x}_{2}\|^{2} \leq \beta-\frac{\gamma}{2}\left[\|\textbf{x}_{1}\|^{2}+\|\textbf{x}_{2}\|^{2}\right]^{2}\leq
\beta+\frac{\gamma^{2}}{4}.\notag
\end{align}
Considering the set
\begin{align}
\mathcal{\widetilde{D}}=\{\textbf{x} \in \mathbb{R}^{2}: \|\textbf{x}\|\leq r\}, \notag
\end{align}
it suffices to have:
\begin{align}
&2r^2\leq -\frac{\gamma}{2}\rightarrow r \leq \sqrt{-\frac{\gamma}{4}} \ ,\notag\\
&r^{4}\leq \beta+\frac{\gamma^{4}}{4}\rightarrow r \leq \sqrt[4]{\beta+\frac{\gamma^{2}}{4}} \ .\notag
\end{align}
Hence,
\begin{align}
r=\min\left(\sqrt{-\frac{\gamma}{4}}\ , \ \sqrt[4]{\beta+\frac{\gamma^{2}}{4}} \ \right), \ \ \ \ \gamma<0,  \ \beta+\frac{\gamma^{2}}{4}>0.
\end{align}
Also, since $\rho=0$, $\xi=\beta+1$ and thus according the LMIs in Corollary 1, since $\lambda< 1$, in order to have $\lambda-\xi>0$, $\beta$ has to be negative. 
As the system is globally one-sided Lipschitz ($\mathcal{D}=\mathbb{R}^{2}$), $\mathcal{D}\cap\mathcal{\widetilde{D}}=\mathcal{\widetilde{D}}$. It is clear that by choosing appropriate values for $\gamma$ and $\beta$, the region $\mathcal{\widetilde{D}}$ can be made arbitrarily large. If we take $\beta=-200, (\xi=-199)$ and $\gamma=-141$, we get $r=5.9372$. Then we take $\alpha=70.6$ (to ensure $\gamma+2\alpha>0$) and solve the LMIs in Corollary 1. We get:
\begin{align}
\lambda=0.999892, \  L=\left[
                   \begin{array}{cc}
                     -1.000000 & 1.000000 \\
                   \end{array}
                 \right]^{T}.
\end{align}
Figure \eqref{Fig2} shows the system trajectories along with their estimates and the system phase plane. This system has a stable limit cycle at $x^{2}+y^{2}=1$ which is inside $\mathcal{\widetilde{D}}$. As this limit cycle is a global attractor, the region $\mathcal{\widetilde{D}}$ is the region of attraction for the observer error dynamics.
For comparison purposes we now consider  the conventional Lipschitz formulation. With $r=5.9372$, the corresponding Lipschitz constant is $l=3r^{2}=105.75$ (compare it with the one-sided Lipschitz constant $\rho=0$). It is highly unlikely that an observer designed based on the conventional Lipschitz approach can work with such a large Lipschitz constant. The \emph{maximum} Lipschitz constant that those observers can handle are normally \emph{at least} an order of magnitude less than this. For example, the maximum admissible Lipschitz constant for the observer designed using \cite{Abbaszadeh1} in this case is $l=1.0324$. The advantage of the new approach in this case, is evident. Note that $A-LC=\left[
                                      \begin{array}{cc}
                                        1 & 0 \\
                                        1 & 0 \\
                                      \end{array}
                                    \right]$ is indeed unstable, confirming our finding of $A-LC$ not being necessarily stable. To verify our result,   lets calculate the Lyapunov derivative. Any positive definite matrix $P$ with condition number $\kappa=\frac{1}{\lambda}$ is acceptable. Lets take $P=\left[
                                      \begin{array}{cc}
                                        \frac{1}{\lambda} & 0 \\
                                        0 & 1 \\
                                      \end{array}
                                    \right]$. From \eqref{ineqVII} we obtain
\begin{align}
\frac{\xi}{\alpha}\lambda_{max}(P)+\lambda_{max}\left[(A-LC)^{T}P+P(A-LC)-\frac{1}{\alpha}P\right]=-0.4187<0.
\end{align}

\section{Conclusions}

This article introduces a new class of nonlinear systems, based on the so-called one-sided Lipschitz continuity condition
commonly used in numerical analysis, as a generalization of the well-known class of Lipschitz systems.
The observer design problem for this class of systems, namely one-sided Lipschitz systems, is established. The advantages of designing observers in this context are explained and the challenges discussed. The mathematical preliminaries of the addressed class of systems are studied and some analysis-oriented tools for the observer stability are provided. Then the challenging design problem is tackled and a solution is proposed based on nonlinear matrix inequalities which are in the cycle casted into the linear matrix inequalities. In order to obtain the proposed solution, a new property for the nonlinear functions, the quadratic inner-boundedness is introduced. Finally, a straightforward observer design procedure is given that can be easily applied to the considered class of system using the available numerically efficient LMI solvers. The efficiency of the approach is shown through an illustrative example.


\bibliographystyle{IEEEtran}
\bibliography{Thesis_References}

\end{document}